\documentclass[aps,prd,showpacs,twoside,superscriptaddress,preprintnumbers,twocolumn,floatfix]{revtex4}
\usepackage[dvips]{graphicx,epsfig,pstricks,rotating}
\usepackage{amsmath}
\usepackage{amssymb}
\newcommand{\beq}{\begin{equation}}
\newcommand{\Phibar}{\bar{\Phi}}
\newcommand{\eeq}{\end{equation}}
\newcommand{\bea}{\begin{eqnarray}}
\newcommand{\eea}{\end{eqnarray}}

\def\slashchar#1{\setbox0=\hbox{$#1$}  % set a box for #1
    \dimen0=\wd0     % and get its size
    \setbox1=\hbox{/} \dimen1=\wd1  % get size of /
    \ifdim\dimen0>\dimen1   % #1 is bigger
       \rlap{\hbox to \dimen0{\hfil/\hfil}} % so center / in box
       #1     % and print #1
    \else     % / is bigger
       \rlap{\hbox to \dimen1{\hfil$#1$\hfil}} % so center #1
       /      % and print /
    \fi}

\begin{document}

\title{Width of the QCD transition in a Polyakov-loop DSE model}

\author{D. Horvati\'c}
%\email{davorh@phy.hr}
\affiliation{Physics Department, Faculty of Science,
University of Zagreb, 
%Bijeni\v{c}ka c. 32, 
Zagreb 10000, Croatia}
\affiliation{Institut f\"ur Physik, Karl-Franzens-Universit\"at Graz, 
%Universit\"atsplatz 5, 
A-8010 Graz, Austria}

\author{D. Blaschke}
%\email{blaschke@ift.uni.wroc.pl}
\affiliation{Institute for Theoretical Physics, University of Wroclaw,
50-204 Wroclaw, Poland}
\affiliation{Bogoliubov  Laboratory of Theoretical Physics, JINR Dubna,
141980  Dubna, Russia}

\author{D. Klabu\v{c}ar}
%\email{klabucar@oberon.phy.hr}
\affiliation{Physics Department, Faculty of Science,
University of Zagreb, 
%Bijeni\v{c}ka c. 32, 
Zagreb 10000, Croatia}

\author{O.~Kaczmarek}
%\email{okacz@physik.uni-bielefeld.de}
\affiliation{Department of Physics, Bielefeld University,
%Postfach 100131
D-33501 Bielefeld, Germany}

\date{\today}
\preprint{BI-TP 2010/49}

\begin{abstract}
We consider the pseudocritical temperatures for the chiral and 
deconfinement transitions within a Polyakov-loop Dyson-Schwinger equation  
approach which employs a nonlocal rank-2 separable model for the effective 
gluon propagator. 
These pseudocritical temperatures differ by a factor of two 
when the quark and gluon sectors are considered separately, but get 
synchronized and become coincident 
%at $T_c=195$ MeV 
when their coupling is switched on.
The coupling of the Polyakov-loop to the chiral quark dynamics narrows 
the temperature region of the QCD transition in which chiral symmetry and 
deconfinement is established. 
We investigate the effect of rescaling
the parameter $T_0$ in the Polyakov-loop potential 
%suggested  by Schaefer, Pawlowski and Wambach 
on the QCD transition for both the logarithmic and polynomial forms of the 
potential.
While the critical temperatures vary in a similar way, the width
of the transition is stronger affected for the logarithmic potential.
For this potential the character of the transition changes from crossover to a 
first order one when $T_0 < 210$ MeV, but it remains crossover in the whole
range of relevant $T_0$ values for the polynomial form.
\end{abstract}

\pacs{11.10.Wx,12.38.Aw,12.38.Mh,12.39.Fe%,25.75.Nq
}

\maketitle

\section{Introduction}
The QCD phase transition between highly excited hadronic matter and the 
quark-gluon plasma is presently under experimental investigation at 
ultra-relativistic 
heavy-ion collider facilities like the Relativistic Heavy-Ion Collider (RHIC)
at the Brookhaven National Laboratory or the Large Hadron Collider (LHC) at 
CERN Geneva.
Its theoretical description requires methods to solve QCD at finite temperature
in the highly nonperturbative low-energy domain.
At present, the only method to obtain ab-initio solutions of QCD in this 
domain is lattice QCD (LQCD).

LQCD calculations are becoming available in the region of 
physical quark masses 
\cite{Aoki:2009sc,Cheng:2009zi,Borsanyi:2010bp,Borsanyi:2010cj,Bazavov:2010sb,Bazavov:2010bx,Bazavov:2010pg,Soldner:2010xk}
and allow for a quantitative description of the equation of state (EoS) and a
determination of the pseudocritical temperature.
Although still afflicted with some uncertainty, those results can be used in 
phenomenological studies of the QCD transition 
relevant for current and future heavy ion experiments at RHIC and LHC.
Hadron gas model calculations give a good description of the EoS
in the confined phase.
Furthermore, statistical models of hadron production  
\cite{BraunMunzinger:2003zd,BraunMunzinger:2001ip} 
can be used to extract the chemical freeze-out temperature which
serves as a lower limit for the deconfinement temperature.

Even so, there are limitations to explore the full QCD phase diagram with LQCD.
Hence dynamical models for the phase structure 
of QCD which can be calibrated with hadron phenomenology and with finite-$T$ 
LQCD remain an important tool.
Particularly useful are chiral quark models of the Nambu--Jona-Lasinio (NJL)-
type \cite{NJL,Volkov:1984kq,Klimt:1989pm,Klevansky:1992qe,Hatsuda:1994pi,Buballa:2003qv}, 
but they suffer from nonphysical quark excitations at low 
temperatures, below $T_c$.
Also the contribution from the gluon sector to the thermodynamics is missing. 
A rather successful generalization has recently been suggested which  
fixes both these problems by coupling the chiral dynamics of the quark sector 
as modeled within the NJL model to a mean-field description of the gluon 
sector with an effective potential as a function of the Polyakov-loop (PL)
variable, fitted to the pure gauge lattice simulations for the Yang-Mills 
pressure \cite{Ratti:2005jh}. 
Within such a PNJL model 
\cite{Meisinger:1995ih,Fukushima:2003fw,Megias:2006bn}, a remarkable agreement 
with other LQCD results as, e.g., for the chiral susceptibilities, could be 
achieved once the temperature in the fit of the gluon mean-field is 
appropriately scaled to the critical temperature obtained in the lattice 
simulations \cite{Ratti:2005jh,Sasaki:2006,Ratti:2007jf}.
The success of this type of chiral quark model in reproducing features of 
lattice QCD simulations has led to a number of applications as well as to 
extensions of the model. 
For example, we would like to mention here the extensions to include 
eight-quark interactions \cite{Sakai:2009dv} and to consider a coupling 
(entanglement) between the scalar coupling constant and the Polyakov-loop
\cite{Sakai:2010rp}.

The deficiency of such a PNJL model is in the poor quark dynamics with, e.g., 
a constant, momentum independent quark mass function. Although the excitation
of quark degrees of freedom is strongly suppressed at low temperatures by 
the destructive interference due the PL phase factors, a dynamical 
confinement mechanism is absent. 
In order to improve the quark dynamics of a PNJL model, the local 
current-current form of the quark interaction has been generalized to a 
nonlocal one  \cite{Blaschke:2007np,Contrera:2007wu,Hell:2008cc}.
In this way, a dynamical quark mass function can be modeled using appropriate
covariant {\it ans\"atze} for the form factor of the nonlocality. 
However, before a quantitative comparison with quark mass functions 
measured in LQCD (see, e.g., Ref.~\cite{Parappilly:2005ei}) can be made, one 
has also to introduce a nonlocal dynamical modification of the Dirac vector 
part of the quark propagator for describing
the quark wave function renormalization as was recently accomplished within
the nonlocal PNJL model in Refs.~\cite{Noguera:2008cm,Contrera:2010kz}.

Alternatively, one can start from QCD Dyson-Schwinger equations 
(DSEs), apply a symmetry-preserving truncation scheme, and solve the 
resulting equations for the Schwinger functions.  In the recent years, 
this approach has reached a new level of maturity (see, e.g., 
Refs.~\cite{Alkofer:2000wg,Roberts:2000aa,Fischer:2006ub} for reviews).
This ongoing progress in DSEs also includes {\it ab initio} type
of calculations \cite{Fischer:2009wc,Fischer:2009gk} pertaining to PL 
and related to other functional approaches.  

There are attempts %, using functional renormalization group methods, 
to derive nonlocal PNJL or PDSE models
%from first principles 
as a low-energy limit of QCD \cite{Kondo:2010ts}, which used insights that have been
obtained in the functional renormalization group method \cite{Braun:2007bx,Braun:2009gm}. 
Functional methods also suggest close relation between the chiral and the 
confinement transition \cite{Braun:2009gm,Braun:2011fw}, and even the relation 
of both of them with the U$_A$(1) symmetry restoration \cite{Alkofer:2011va}. 
These studies are worth a systematic further development.

Since QCD applications at $T>0$ are in general considerably 
more difficult than at $T=0$, it is often useful to take a more 
phenomenological approach: one chooses a suitable  (e.g., 
phenomenologically successful) effective dressed-gluon propagator model
and solves the resulting $T>0$ DSEs for the Schwinger functions 
in the quark sector \cite{Roberts:2000aa}. 
%start from the finite-temperature generalization of the
%QCD Dyson-Schwinger equations (DSE), apply a symmetry-preserving truncation
%scheme and solve the resulting equations for the Schwinger functions in the 
%quark sector for suitably chosen gluon propagator models \cite{Roberts:2000aa}.
Such an approach has many advantages over NJL-type models as, e.g., a dynamical
confinement mechanism.  
However, as has been analyzed in \cite{Blank:2010bz} the critical temperatures 
obtained within such non-perturbative low-energy QCD models turn out to be too 
low when compared with LQCD results.   

In the present work we suggest that this shortcoming might have its origin in 
residual color correlations and we investigate a generalization where the PL 
is coupled to the quark sector DSE (PDSE).
For the purpose of this exploratory approach, we will restrict ourselves in 
this PDSE model to a rank-2 separable form of the effective gluon propagator 
\cite{Blaschke:2000gd}
which at the rainbow-ladder level of truncation is equivalent 
to a full DSE model with translation-invariant gluon propagator 
once the model form factors of the interaction are chosen appropriately
\cite{Burden:1996nh}. 
%{\bf Ongoing progress ... \cite{Fischer:2009wc,Fischer:2009gk}}
%
%{\bf There are attempts, using functional renormalization group method, to 
%derive nonlocal PNJL or PDSE models 
%%from first principles 
%as a low-energy limit of QCD, which are worth a systematic further
%development \cite{Kondo:2010ts}, which use insight bering obtained in
%functional renormalization group method \cite{Braun:2007bx,Braun:2009gm}. }
%Functional methods also suggest close relation between the chiral and 
%the confinement transition \cite{Braun:2009gm}, and even the relation of 
%both of them with the U$_A$(1) symmetry restoration \cite{Alkofer:2011va}.
For the time being one may take the pragmatic approach of 
modeling form factors of the rank-2 separable interaction such that both, the
dynamical quark mass function and the wave function renormalization as 
measured in LQCD at zero temperature can be reproduced to high accuracy.
With an interaction model fixed in this way, the approach developed here can
be used to predict thermodynamics and hadron properties of low-energy QCD
at finite temperature, calibrate the results with LQCD and extend the approach
subsequently to the whole QCD phase diagram, i.e., into regions presently 
inaccessible to LQCD.   

In the present work, by analyzing the temperature behavior of the order 
parameters in the model, 
the dynamically generated light and strange quark masses as well as the 
PL variable, we find that the critical temperatures for the chiral and the 
deconfinement transitions measured by the peaks in the corresponding 
susceptibilities (defined here as the temperature derivatives of the order 
parameters) coincide at the per mille level of accuracy. 
We will discuss the effect of rescaling the critical temperature parameter 
$T_0$ of the PL potential 
\cite{Braun:2005uj,Schaefer:2007pw,Schaefer:2009ui,Herbst:2010rf} 
once applications with a finite 
number of quark flavors and a possible chemical potential are considered and
how this affects the width of the QCD transition region. 

As a possible application of this class of models we discuss the 
temperature dependence of scalar and pseudoscalar meson properties at finite 
temperatures towards the chiral symmetry restoration.
We devote special attention to the investigation of a 
Gell-Mann--Oakes--Renner (GMOR)- like relationship for the mass of 
pseudoscalar mesons at finite temperatures.
We find that the GMOR-like relation proves to be robust up to the critical 
temperature, as a manifestation of the implemented confinement mechanism at the
rainbow-ladder level of the description, where the coupling to the PL provides 
a very effective suppression of the quark degrees of freedom in the medium
which otherwise induce a medium dependence which is unphysical below $T_c$.
In this regime, medium effects have to be absent as long as hadronic 
excitations in the medium are disregarded.
%We thus want to underline as a result of the present work the importance of
%going beyond the rainbow-ladder level for discussing the nature of the QCD 
%transition and the behavior of chiral and deconfinement order parameters in 
%its vicinity. 

\section{Separable PDSE model}
\subsection{Thermodynamical potential and order parameters}
The central quantity for the analysis of the thermodynamical behavior is the
thermodynamical potential which in the PDSE approach is a straightforward 
generalization of the standard CJT functional 
\cite{Cornwall:1974vz,Haymaker:1991} 
\begin{widetext}
\begin{eqnarray}
\label{Om_PL-DSE}
\Omega(T)&=&- T \ln Z(T)
=\mathcal{U}(\Phi,\bar{\Phi})
-T {\rm Tr}_{\vec{p},n,\alpha,f,D} 
\left[ \ln \{S_f^{-1}(p_n^\alpha,T)\} 
- \Sigma_f(p_n^\alpha,T)\cdot S_f(p_n^\alpha,T)\right]~,
\end{eqnarray}
\end{widetext}
where the full quark propagator for the flavor $f=u,d,s$,
\begin{eqnarray}
S_f^{-1}(p_n^\alpha,T)
&=& S_{f,0}^{-1}(p_n^\alpha,T)+\Sigma_f(p_n^\alpha,T)\nonumber\\ 
&=& i\vec{\gamma} \cdot \vec{p}\; A_f((p_n^\alpha)^2,T)\nonumber\\ 
  &&  + i \gamma_4 \omega_n\; C_f((p_n^\alpha)^2,T)+ B_f((p_n^\alpha)^2,T),
\end{eqnarray}
is defined by the DSE for the quark self-energy $\Sigma$, see below.
The Polyakov-loop potential is  first 
taken in the form \cite{Roessner:2006xn}
\begin{widetext}
\begin{equation}
\frac{\mathcal{U}_\mathrm{log}(\Phi,\bar{\Phi})}{T^4}=-\frac{1}{2}a(T)\bar{\Phi}\Phi+
b(T)\ln\left[1-6\bar{\Phi}\Phi+4({\bar{\Phi}}^3+\Phi^3)-3(\bar{\Phi}\Phi)^2
\right]
\label{logU}
\end{equation}
\end{widetext}
with
%\begin{equation}
$a(T)=a_0+a_1(T_0/T)+a_2(T_0/T)^2$~,
$b(T)=b_3(T_0/T)^3$~.
%\end{equation}
The corresponding parameters are taken from Ref.~\cite{Roessner:2006xn}, 
$a_0=3.51$, $a_1=-2.47$, $a_2=15.22$ and $b_3=-1.75$ where they have been 
adjusted to fit the pressure obtained in lattice gauge theory simulations
of SU(3) Yang-Mills theory. In most of the literature on the PNJL model,
the parameter $T_0=270$ MeV has been taken over for applications in QCD with
$N_f$ quark flavors while following 
Ref.~\cite{Schaefer:2007pw} its 
dependence on quark flavors and chemical potential should be invoked.
Accordingly, for the case $N_f=2+1$ discussed in the present work, in
\cite{Schaefer:2007pw} the value $T_0=187$ MeV is suggested with an error 
margin of about 30 MeV.
Applying the Matsubara formalism of finite temperature field theory, the 
squared quark 4-momenta are to be replaced by 
$(p^\alpha_n)^2=(\omega^\alpha_n)^2 + \vec{p}^{\,2}$,
$\omega^\alpha_n=\omega_n+\alpha\phi_3$,  where 
$\omega_n=(2n+1)\pi T$ are the fermionic Matsubara frequencies and the 
indices  $\alpha=-1,0,+1$ specify the three quark colors and their coupling 
to the parameter $\phi_3$ of the temporal gauge field. 
At vanishing chemical potential, the Polyakov loop is given by
\begin{eqnarray}
\Phi=\bar{\Phi}&=&
\frac{1}{N_c}\left(1+e^{i\frac{\phi_3}{T}}+e^{-i\frac{\phi_3}{T}}\right)
\nonumber\\
&=&\frac{1}{N_c}\left(1+2\cos\left(\frac{\phi_3}{T}\right)\right)~.
\end{eqnarray}

In order to check the sensitivity to various parameterizations of
the Polyakov-loop potential, we will also try the polynomial
form \cite{Ratti:2005jh}
\begin{widetext}
\begin{equation}
\frac{\mathcal{U}_\mathrm{poly}(\Phi,\bar{\Phi})}{T^4} = -\frac{b_2(T)}{4}
\left(|\Phi|^2+|\bar\Phi|^2 \right) 
-\frac{b_3}{6}(\Phi^3+\Phibar^3)+\frac{b_4}{16}
\left(|\Phi|^2+|\Phibar|^2\right)^2
\label{polyU}
\end{equation}
\end{widetext}
with the temperature-dependent coefficient
\begin{equation}
  b_2(T) = a_0 + a_1 \left(\frac{T_0}{T}\right) + a_2
  \left(\frac{T_0}{T}\right)^2 + a_3 \left(\frac{T_0}{T}\right)^3 
\end{equation}
and the the set of parameters from Ref.~\cite{Ratti:2005jh},
$a_0=6.75$, $a_1=-1.95$, $a_2=2.625$, $a_3 = -7.44$, $b_3=0.75$, and 
$b_4=7.5$.

For the effective gluon propagator in a Feynman-like gauge, 
$g^2 D_{\mu\nu}^{\rm eff} (p-q) = \delta_{\mu\nu} D(p^2,q^2,p\cdot q)$,  
we employ a rank-2 separable ansatz \cite{Blaschke:2000gd}
\begin{eqnarray}
D(p^2,q^2,p\cdot q)&=&D_0 {\cal F}_0(p^2) {\cal F}_0(q^2)\nonumber\\ 
&&+ D_1 {\cal F}_1(p^2) (p\cdot q ) {\cal F}_1(q^2)~,
\end{eqnarray}
so that the quark propagator amplitudes are given by 
\begin{eqnarray}
B_f((p_n^\alpha)^2,T) &=& {m}_f^0 + b_f(T) {\cal F}_0((p_n^\alpha)^2)~,\\
A_f((p_n^\alpha)^2,T) &=& 1+ a_f(T) {\cal F}_1((p_n^\alpha)^2)~,\\ 
C_f((p_n^\alpha)^2,T) &=& 1+ c_f(T) {\cal F}_1((p_n^\alpha)^2)~,
\end{eqnarray}
and their analytic properties are defined by the choice of the form factors.
In the present work we will use the functions 
\cite{Blaschke:2007ce,Horvatic:2007wu}
\begin{eqnarray}
{\cal F}_0(p^2)&=&\exp(-p^2/\Lambda_0^2)~,\\
{\cal F}_1(p^2)&=&\frac{1+\exp(-p_0^2/\Lambda_1^2)}
{1+\exp((p^2-p_0^2)/\Lambda_1^2)}~,
\end{eqnarray}
which satisfy the constraints ${\cal F}_0(0)={\cal F}_1(0)=1$ and 
${\cal F}_0(\infty)={\cal F}_1(\infty)=0$. 
Their functional form can be chosen such that the 4-momentum dependence of the 
dynamical mass function $M(p)=B(p)/A(p)$ and the the wave function 
renormalization $Z(p)=1/A(p)$ is in good agreement 
\cite{Noguera:2008cm}
with LQCD simulations of the quark propagator \cite{Parappilly:2005ei}.
Models which employ a rank-1 separable ansatz (see, e.g., 
\cite{Blaschke:2007np,Hell:2008cc})
result in $A(p)=Z(p)=1$ and miss an important aspect of quark dynamics in QCD.

The temperature-dependent gap functions $a_f(T)$, $b_f(T)$ and $c_f(T)$ are
obtained as solutions of the DSE for the quark self energy in rainbow-ladder
truncation as \cite{Blaschke:2000gd}
\begin{widetext}
\begin{eqnarray}
a_f(T) = \frac{8 D_1}{27}\,  T \sum_{n,\alpha} \int \frac{d^3p}{(2\pi)^3}\,{\cal F}_1((p_n^\alpha)^2)\, \vec{p}^{\,2}\, A_f((p_n^\alpha)^2,T)\; 
d_f^{-1}((p_n^\alpha)^2,T) \; ,
\label{oldgap1}\\
 c_f(T) = \frac{8 D_1}{9}\,  T \sum_{n,\alpha} \int \frac{d^3p}{(2\pi)^3}\,{\cal F}_1((p_n^\alpha)^2)\, (\omega^\alpha_n)^2\, C_f((p_n^\alpha)^2,T)\; 
d_f^{-1}((p_n^\alpha)^2,T) \; ,
\label{oldgap2}\\
 b_f(T) = \frac{16 D_0}{9}\,  T \sum_{n,\alpha} \int \frac{d^3p}{(2\pi)^3}\,{\cal F}_0((p_n^\alpha)^2)\, B_f((p_n^\alpha)^2,T)\; 
d_f^{-1}((p_n^\alpha)^2,T) \; ,
\label{oldgap3}
\end{eqnarray}
\end{widetext}
where the denominator function is
%\begin{equation}
$d_f((p_n^\alpha)^2,T) = \vec{p}^{\,2}A_f^2((p_n^\alpha)^2,T)+(\omega^\alpha_n)^2C_q^2((p_n^\alpha)^2,T)+ B_q^2((p_n^\alpha)^2,T).
$
%\end{equation}
Eqs.~(\ref{oldgap1})-(\ref{oldgap3}) correspond to minima of the 
thermodynamical
potential (\ref{Om_PL-DSE}) with respect to a variation of the temperature 
dependent gap functions $a_f(T),~b_f(T),~c_f(T)$ and have to be supplemented by
a corresponding gap equation for the Polyakov loop which follows from the 
extremum condition $\partial \Omega/\partial \Phi|_{\rm min}=0$.
Once the gap equations are solved for different temperatures, one can extract
the pseudocritical temperatures for chiral and deconfinement transitions from 
the peak positions of the temperature derivatives of the corresponding order 
parameters, the quark mass functions
$m_f(T)=[m_f^0+b(T)]/[1+a_f(T)]$ and the Polyakov loop $\Phi(T)$, 
respectively. 
For a discussion of the quark mass function as an order parameter of the chiral
transition see, e.g., Refs.~\cite{Bender:1996bm,Blaschke:1998mp,Holl:1998qs}.

\subsection{Pion, kaon and sigma meson at finite temperature}

At \mbox{$T=0$} the mass-shell condition for a meson as a $q \bar q'$ 
bound state of the
Bethe-Salpeter equation (BSE)
is equivalent to the appearance of a pole in the $q \bar q'$ scattering 
amplitude as a
function of $P^2$. The $q\bar q'$ meson Bethe-Salpeter bound-state 
vertex
$\Gamma_{f f'}(p,P)$ is the solution of the BSE
\begin{widetext}
\begin{eqnarray}
\label{bse}
-\lambda(P^2)\Gamma_{f f'}(p,P) = \frac{4}{3} \int 
\frac{d^4\ell}{(2\pi)^4}
  g^2  D_{\mu\nu}^{\mathrm{eff}} (p-\ell)
\gamma_\mu S_f(\ell_+) \Gamma_{f f'}(\ell,P) S_{f'}(\ell_-) \gamma_\nu, \,
\label{BSE}
\end{eqnarray}
\end{widetext}
where the index $f$ (or $f'$) stands for
the quark (or antiquark) flavor ($u, d$ or $s$),
$P$ is the total 4-momentum, and $\ell_{\pm}=\ell\pm P/2$.
The meson mass is identified from $\lambda(P^2=-M^2)=1$.

For example, with the separable interaction, the allowed form of the 
solution of Eq. (\ref{bse}) for the pseudoscalar Bethe-Salpeter amplitude is

\begin{equation}
\Gamma_{P}(\ell;P) = \gamma_5 \left(i E_{P} (P^2) +
                      \slashchar{P} F_{P}(P^2)\right) \; {\cal F}_0(\ell^2).
                      \label{psbsa}
\end{equation}                    

For scalar mesons we will use a truncated form of the Bethe-Salpeter amplitude 
(i.e. we take only the dominant contribution)

\begin{equation}
\Gamma_{S}(\ell;P) =  E_{S}(P^2) \; {\cal F}_0(\ell^2).      
\end{equation}

At $T\neq0$ in the Matsubara formalism, the $O(4)$ symmetry is broken
by the heat bath and we have \mbox{$P \to P_m=(\nu_m,\vec{P})$} where 
\mbox{$\nu_m = 2m \pi T$}.  
Bound states and the poles they generate in propagators may be investigated
through polarization tensors, correlators, or Bethe-Salpeter eigenvalues.  
This pole structure is characterized by information at discrete points 
$\nu_m$ on the imaginary energy axis and at a continuum of 3-momenta. 
One may search for poles as a function of $\vec{P}^2$ thus identifying the 
so-called spatial or screening masses for each Matsubara mode.  
These serve as one particular characterization of the propagator 
and the \mbox{$T > 0 $} bound states. 
In the present context, the eigenvalues of the BSE become $\lambda(P^2) \to
\tilde{\lambda}(\nu_m^2,\vec{P}^2;T)$. 
The spatial screening masses are identified by zeros of 
$1-\tilde{\lambda}(0,\vec{P}^2;T)$.

The general form of the finite-$T$ pseudoscalar and scalar Bethe-Salpeter amplitude 
allowed by the separable model for the lowest Matsubara mode $\nu_0 =0$ 
(as required for the spatial meson modes of interest here) is
\begin{eqnarray}
\Gamma_{P}(\ell_n^\alpha;\vec{P}) &=&\gamma_5 \left(i E_{P} (\vec{P}^2)
+  \vec{\gamma} \cdot \vec{P} F_{P}(\vec{P}^2)\right) \; {\cal 
F}_0((\ell_n^\alpha)^2) \; \nonumber\\
\Gamma_{S}(\ell_n^\alpha;\vec{P}) &=&  E_{S}(\vec{P}^2) \; {\cal F}_0((\ell_n^\alpha)^2)
\label{pibsaT}
\end{eqnarray}
One can then write the BSE for the spatial masses as
\begin{widetext}
\begin{equation}
\tilde{\lambda}(0,\vec{P}^2;T)\Gamma_{f f'}(p_m^\alpha;\vec{P}) 
= \frac{4}{9}T \sum_{n,\alpha} \int \frac{d^3\ell}{(2\pi)^3}
  g^2  D_{\mu\nu}^{\mathrm{eff}} 
(\omega_m^\alpha-\omega_n^\alpha,\vec{p}-\vec{\ell})
\gamma_\mu S_f((\ell_n^\alpha)_+) \Gamma_{f
f'}(\ell_n^\alpha,\vec{P}) S_{f'}((\ell_n^\alpha)_-) \gamma_\nu.
\end{equation}
\end{widetext}

For example, BSE for the scalar $\sigma$ meson is
\begin{widetext}
\begin{eqnarray}
\tilde{\lambda}_S(0,\vec{P}^2;T) &=& 
\frac{16 D_0}{9} T \sum_{n,\alpha} \int \frac{d^3\ell}{(2\pi)^3} {\cal F}_0^2((\ell_n^\alpha)^2) 
\left[(\omega_n^\alpha)^2\sigma_{C,u}((\ell_n^\alpha)_+)\sigma_{C,u}((\ell_n^\alpha)_-)\right. 
 \\ \nonumber
&+& \left.\left(\vec{\ell\;}^2-\frac{\vec{P}^2}{4}\right)\sigma_{A,u}((\ell_n^\alpha)_+)\sigma_{A,u}((\ell_n^\alpha)_-)
                                       - \sigma_{B,u}((\ell_n^\alpha)_+)\sigma_{B,u}((\ell_n^\alpha)_-) \right]~,
\end{eqnarray}
\end{widetext}
where \mbox{$\sigma_{A,f} = A_f/d_f$}, \mbox{$\sigma_{C,f} = C_f/d_f$} and
\mbox{$\sigma_{B,f} = B_f/d_f$}.
Further details on the analysis of mesonic spatial screening masses in other 
channels and for specific model interaction kernels can be found, e.g., in 
Ref.~\cite{Blaschke:2000gd,Blaschke:2007ce,Horvatic:2007wu}. 
Here, we have generalized this approach by accounting for the PL phase factors 
entering the quark propagators and the interaction kernels.

\section{Results and Discussion}

For the numerical calculations we fix the free parameters of the model 
at $T=0$ as in Refs.~\cite{Kalinovsky:2005kx,Horvatic:2007wu,Blaschke:2007ce},
to reproduce in particular the vacuum masses of the 
pseudoscalar and vector mesons, $M_\pi=140$ MeV, $M_K=495$ MeV, $M_\rho=770$ MeV, the pion decay 
constant $f_\pi=92$ MeV, and decay widths, $\Gamma_{\rho^0\to\mathrm{e}^+\mathrm{e}^-}=6.77$ keV, 
$\Gamma_{\rho\to\pi\pi}=151$ MeV as basic requirements from low-energy QCD phenomenology. 
   
For clarity we point out that at $T=0$, the model without PL coincides with the otherwise 
same model including PL (just as they do in the NJL vs. PNJL case). We thus obtain the same 
parameter set as in Refs.~\cite{Kalinovsky:2005kx,Horvatic:2007wu,Blaschke:2007ce},
namely $m_u^0=m_d^0=m_q^0=5.49$~MeV, 
$m_s^0=115$~MeV, $D_0\Lambda_0^2 = 219$, $D_1\Lambda_0^4 =  69$,
$\Lambda_0=0.758$~{GeV}, $\Lambda_1=0.961$~GeV and $p_0=0.6$~{GeV}.

\begin{figure}[!ht]
\includegraphics[width=.45\textwidth,angle=0]{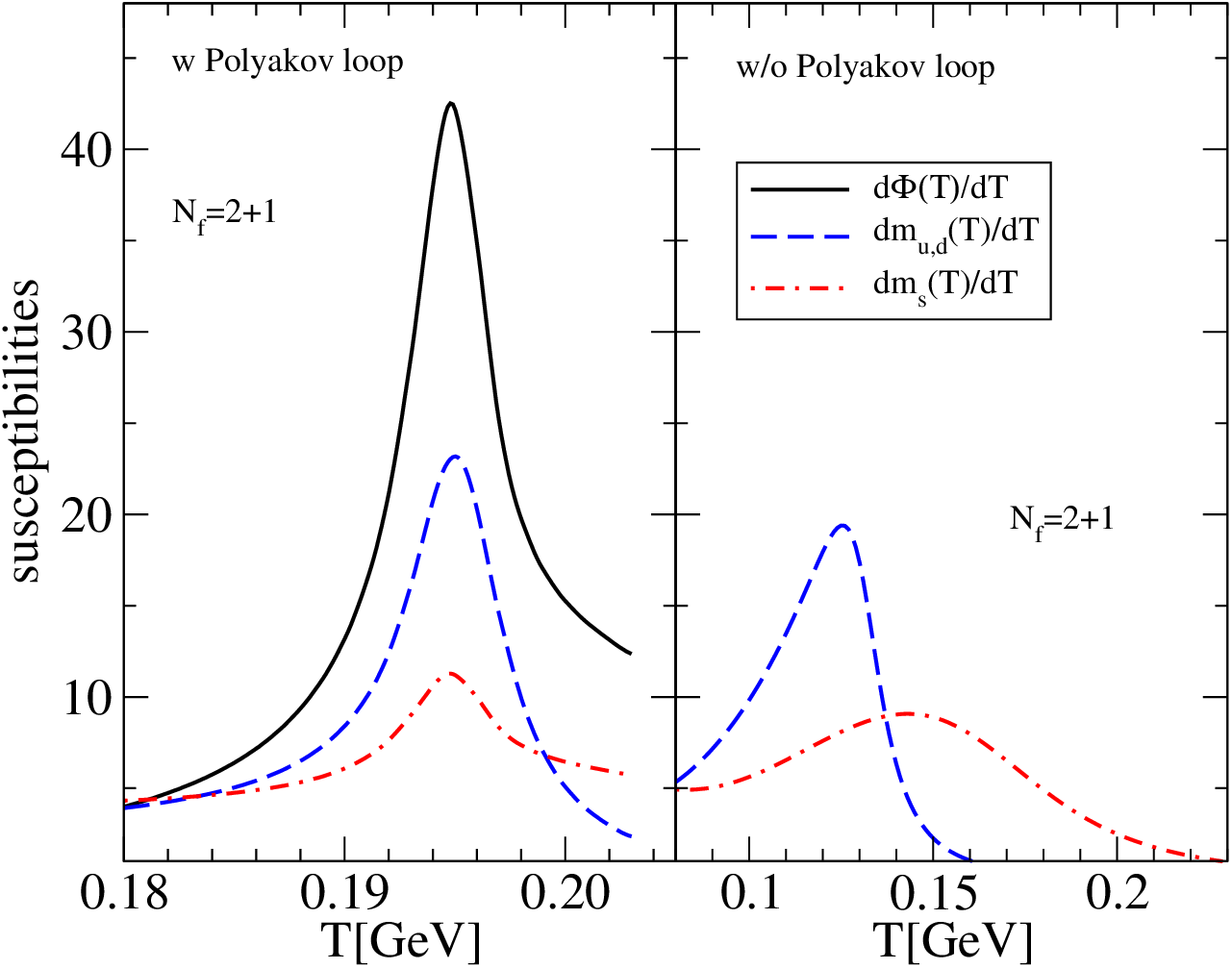}
\caption{(Color online)
Quark mass susceptibilities (blue dashed line: light flavors; red dash-dotted 
line: strange flavor) with coupling to the Polyakov loop (left panel) and
without it (right panel) as a function of the temperature. 
Note that without coupling to the Polyakov loop the chiral transition 
temperature is unrealistically low and the peak value for the light flavors 
is different from that for the strange one.
}
\label{suscpNoPL}
\end{figure}

\subsection{Order parameters for chiral and deconfinement transition}

In Fig.~\ref{suscpNoPL} we show the resulting temperature dependence of the 
derivatives of the quark mass being an order parameter of the chiral phase 
transition and of the Polyakov loop expectation value as an order parameter
of the deconfinement transition. The peak values are attained at the 
corresponding pseudocritical temperatures for the chiral ($T_\chi$) and 
deconfinement ($T_d$) transitions, respectively.
In the right panel of Fig.~\ref{suscpNoPL} we show the results when the quark 
and gluon sectors are uncoupled. 
In this case we have in the light quark sector $T_\chi=128$ MeV, whereas 
$T_d=270$ MeV according to the parametrization of the PL potential in the 
pure gauge sector.
The value obtained for $T_\chi$ is in the typical range found in the DSE 
approach \cite{Blank:2010bz}.
The peak position of the chiral susceptibility in the strange quark sector 
does not coincide with the one in the light quark sector in this case.
When the quark and gluon sectors are coupled these temperatures get
synchronized so that $T_c = T_\chi= T_d = 195$ MeV, as is demonstrated in 
the left panel of Fig.~\ref{suscpNoPL}.

 \begin{figure}[ht]
\includegraphics[width=.45\textwidth,angle=0]{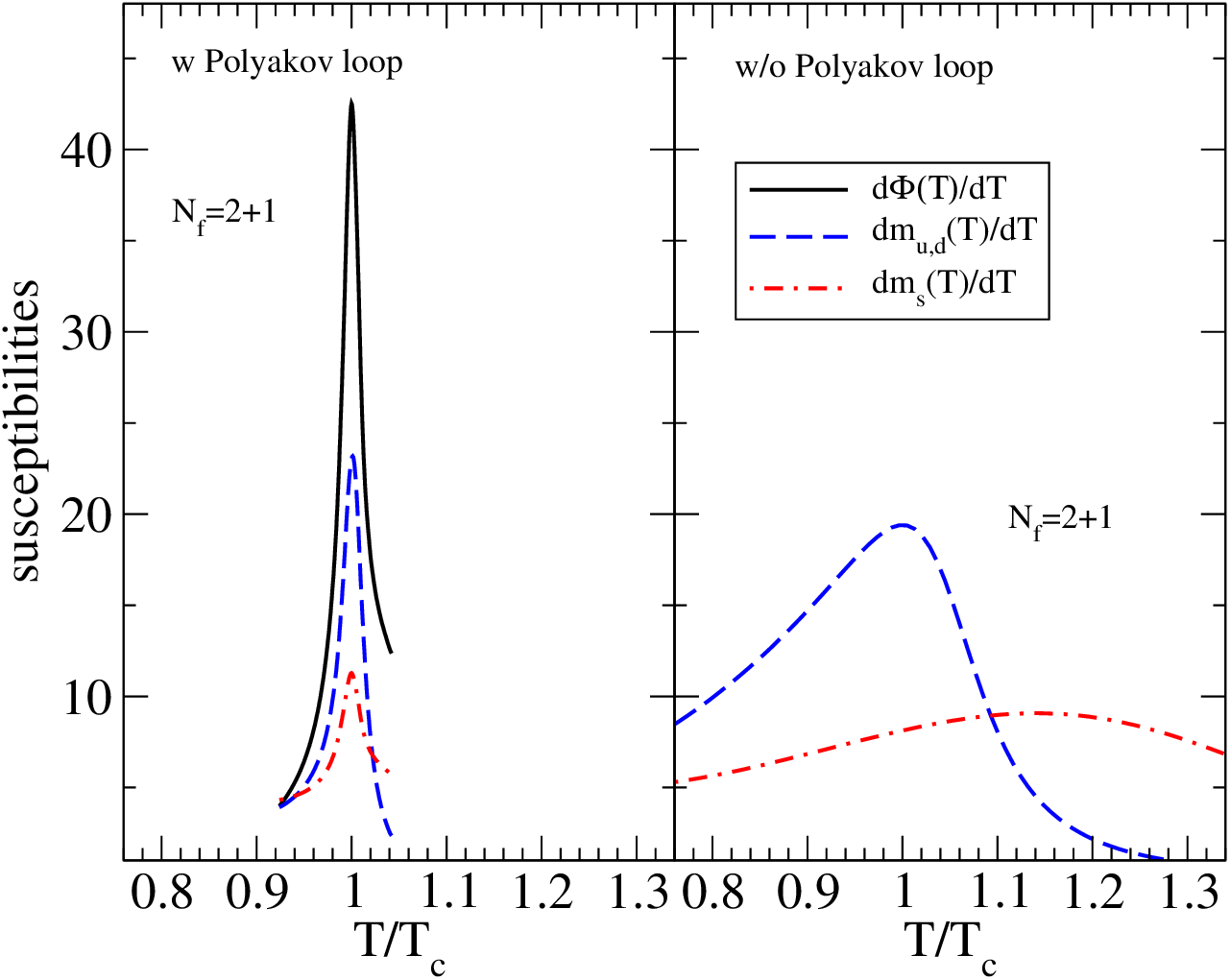}
\caption{(Color online)
Same as Fig.~\ref{suscpNoPL}, but as a function of the scaled temperature
$T/T_c$ with $T_c=195$ MeV (left panel) and  $T_c=T_\chi=128$ MeV 
(right panel). Without coupling to the Polyakov-loop $T_d=2.11~T_c$ is outside 
the range shown. 
}
\label{suscpNoPL-r}
\end{figure}

At the same time, when coupling the PL potential to the chiral quark sector,
the width of the transition region collapses to a tiny temperature interval
around $T_c$, as is demonstrated in Fig.~\ref{suscpNoPL-r} where the 
susceptibilities are shown as functions of the scaled temperature $T/T_c$ in
the same interval with (left panel) and without (right panel) coupling the 
quark sector to the Polyakov loop potential.

Both effects of coupling the chiral quark sector to the PL, the synchronization
of the chiral and deconfinement transitions as well as the narrowing of the 
width of the QCD transition region, are obtained in a similar way for the 
polynomial PL potential (\ref{polyU}).

\begin{figure}[ht]
\includegraphics[width=.45\textwidth,angle=0]{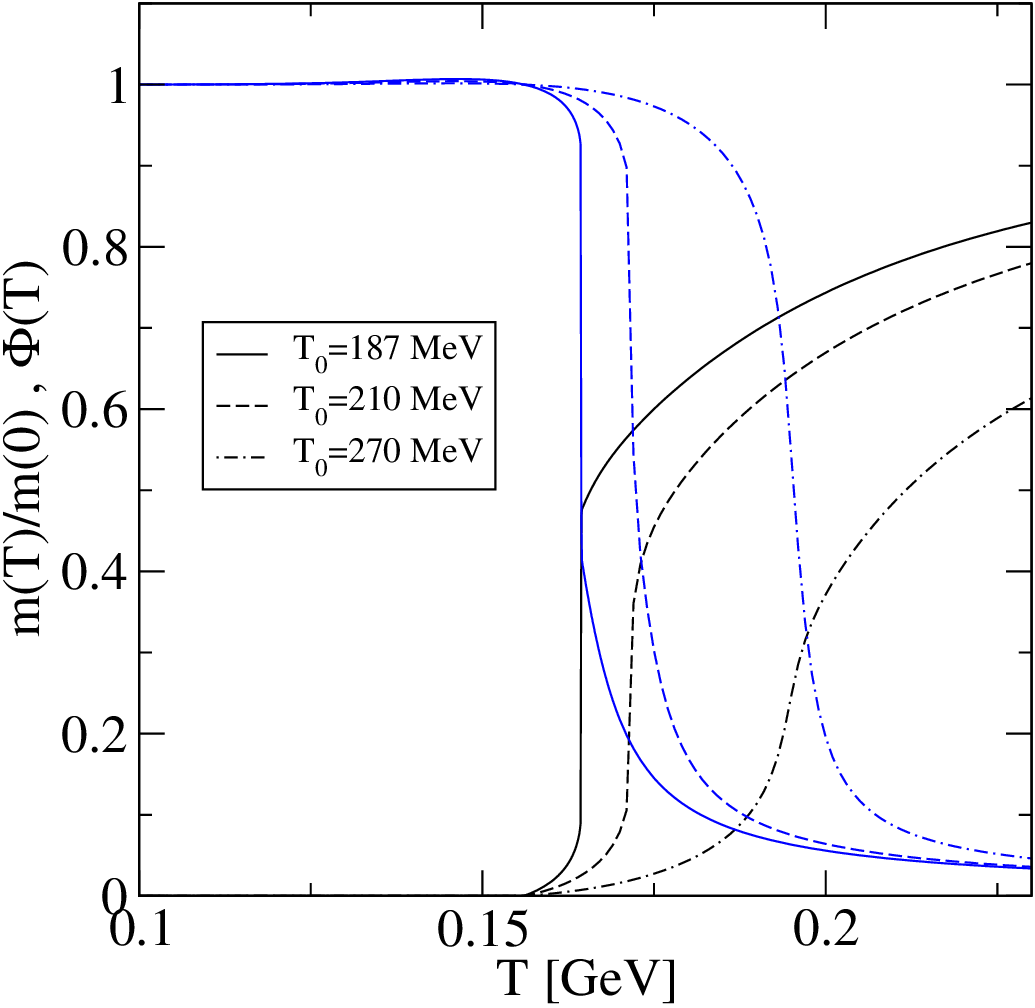}
\caption{(Color online)
Temperature dependence of the order parameters for chiral symmetry breaking 
($m(T)/m(0)$, blue lines) and for deconfinement ($\Phi(T)$, black lines) for 
different choices for the parameter $T_0$ in the Polyakov-loop potential.
}
\label{mT-PhiT}
\end{figure}

The value obtained for the QCD transition temperature, $T_c=195$ MeV 
 (193 MeV) for the logarithmic (polynomial) PL potential, is
closer to recent LQCD results than the one obtained in PNJL or rank-1 
separable nonlocal PNJL models but unsatisfactory for a quantitative 
description. 
Within the framework of the PQM model, it has been suggested 
\cite{Schaefer:2007pw} to rescale the $T_0$ parameter of the PL potential depending on the quark
flavor content of the system and the chemical potential. 
We will follow such a prescription also in the present approach.  
In Fig.~\ref{mT-PhiT}, we show the resulting temperature dependence of the 
order parameters for chiral symmetry breaking (the normalized mass function 
$m(T)/m(0)$) and for deconfinement (the PL $\Phi(T)$) for three values of 
$T_0$. According to \cite{Schaefer:2007pw} the case $T_0=187$ MeV corresponds
to $N_f=2+1$ while $T_0=270$ MeV is the value for the pure gauge theory where
the deconfinement is a first order phase transition. 
The coupling to the chiral quark dynamics changes the character of this 
transition to a crossover. Lowering the $T_0$ parameter to 187 MeV
changes both deconfinement and chiral restoration to strong first order
phase transitions! This change of character happens at the critical value 
$T_0=210$ MeV, also shown in Fig.~\ref{mT-PhiT}.

\begin{figure}[ht]
\includegraphics[width=.45\textwidth,angle=0]{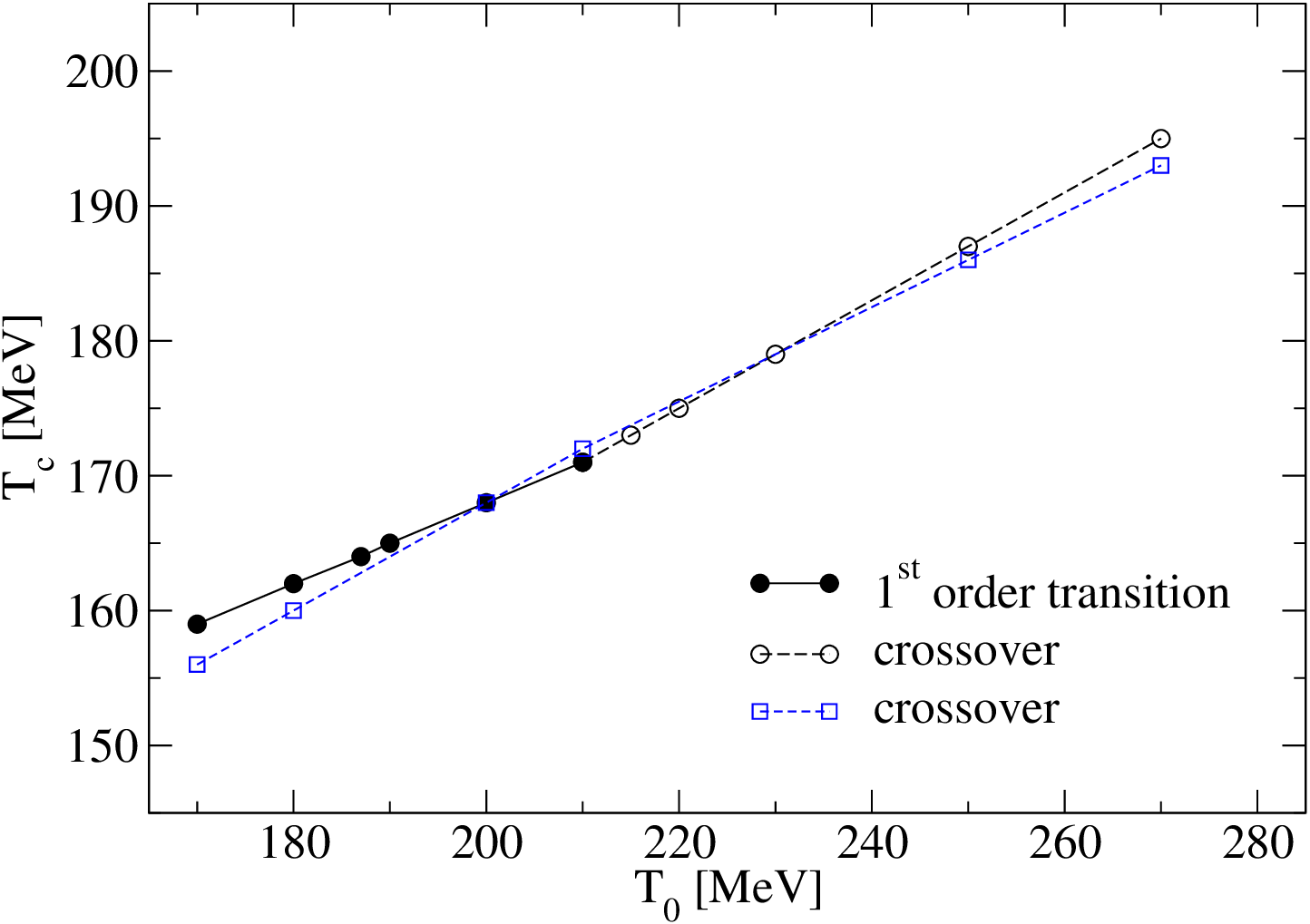}
\caption{
(Color online) Pseudocritical temperature for the chiral restoration 
transition vs. parameter $T_0$ of the Polyakov-loop potential in
the logarithmic form (\ref{logU}) (black circles) and in the 
polynomial form (\ref{polyU}) (blue squares). 
For further details, see text.
}
\label{TcT0}
\end{figure}

In Fig.~\ref{TcT0}, we summarize this finding by showing the dependence of 
$T_c$ on the $T_0$ parameter of the PL potentials (\ref{logU}) and 
(\ref{polyU}). For the logarithmic PL potential (\ref{logU}), the 
positions of first order transitions are characterized by the full dots 
connected by a solid line, while the crossover transitions are given as 
open dots connected by a dashed line.  
Two regions of linear dependence can be identified when using the logarithmic 
PL potential: 
$T_c={\rm const} + 0.30~T_0$ for $T_0<210$ MeV and 
$T_c={\rm const} + 0.40~T_0$ for $T_0>210$ MeV. 
When using the polynomial form (\ref{polyU}) of the PL potential, we find the
linear dependence as $T_c={\rm const} + 0.36~T_0$. 
The  change in the character of the QCD transition from a crossover for 
$T_0>210$ MeV to a first oder transition for $T_0<210$ MeV is accompanied by 
a sudden change in slope at $T_0=210$ MeV. 
It is remarkable that the $T_0$- rescaling introduced to account for a quark 
flavor dependence of the PL potential when applied to the nonlocal separable 
PDSE model considered here, results in an obvious contradiction with 
LQCD concerning the character of the QCD transition: 
while in LQCD for $N_f=2+1$ the finite-T transition is a crossover
\cite{Aoki:2006we,Ejiri:2009ac}, 
the application of the suggested reparametrization with the corresponding 
value $T_0=187$ MeV leads in the present model to a first order transition. 

On the other hand, for the polynomial PL potential (\ref{polyU})
the transition is a crossover for any of the considered values of $T_0$. 
Fig.~\ref{TcT0} illustrates this by the dashed line connecting the
points depicted by squares.

\subsection{Meson screening masses at finite $T$}

\begin{figure}[ht]
\includegraphics[width=.45\textwidth,angle=0]{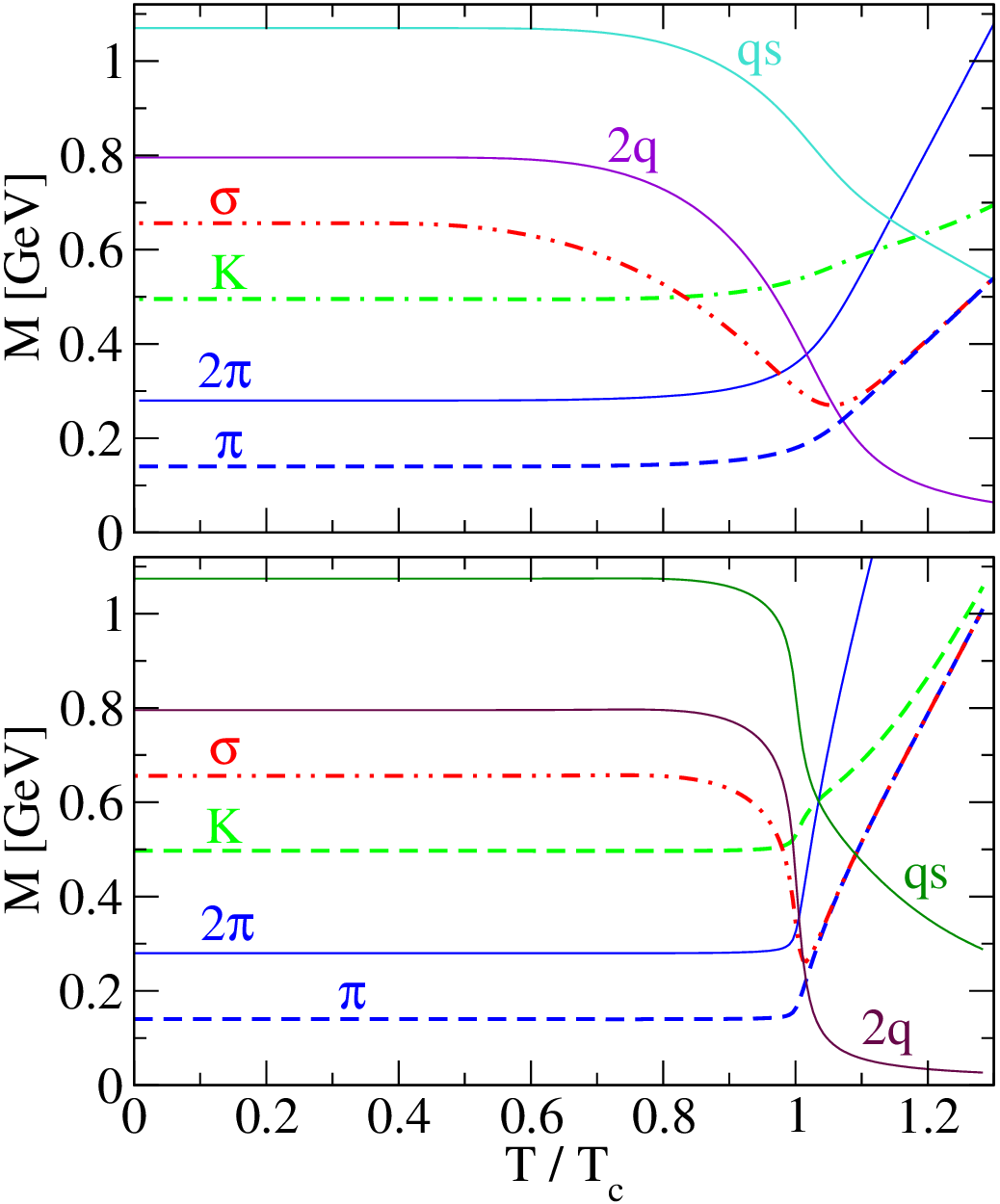}
\caption{
Temperature dependence of pseudoscalar ($\pi$, K) and scalar ($\sigma$) meson 
masses in the present model without (upper panel) and with (lower panel) 
coupling to the Polyakov-loop potential. The temperature $T_{\sigma-2\pi}$ 
($T_\sigma-{\rm 2q}$) where the $\sigma$ meson mass equals the double pion 
mass (double quark mass) determines the threshold for closing (opening) the 
$\sigma \to 2 \pi$ ($\sigma \to 2 q$) decay channel.
}
\label{TcMps}
\end{figure}

Following the approach to spatial meson screening masses developed in 
\cite{Blaschke:2000gd} in its generalization by the coupling to the Polyakov
loop as given above, we have evaluated the temperature dependence of 
scalar and pseudoscalar meson masses.
In Fig.~ \ref{TcMps}, we show the results with (lower panel) and without
(upper panel) PL coupling together with the behavior of the threshold to
the continuum estimated by the sum of the corresponding quark mass functions.
Since the PL coupling leads to a strong suppression of thermal quark 
excitations below the critical temperature, the continuum thresholds are
almost constant with a sudden drop in the vicinity of $T_c$. 
This behavior is reflected in the temperature dependence of the meson 
screening masses. 
As a quantitative measure for the width of the QCD transition, we suggest 
to consider either the difference of $\sigma$- and $\pi$- mass squared,
$\Delta_{\sigma-\pi}^2=M_\sigma^2 - M_\pi^2$, or the difference between the 
temperature $T_{\sigma-2\pi}$ where the $\sigma$ meson mass equals the double 
pion mass (the threshold for closing the $\sigma \to 2 \pi$ decay channel
and the temperature $T_\sigma-{\rm 2q}$ where the $\sigma$ meson mass equals 
the double quark mass (the threshold for opening the $\sigma \to 2 q$) 
decay channel). 
Both measures reveal that the width of the transition region reduces from 
about $10~\%$ without to about $1~\%$ with PL coupling. 
Note that our results for the temperature dependence of the $\sigma$- and 
$\pi$- meson without PL coupling are very similar to those obtained earlier
within the DSE approach \cite{Maris:2000ig}.

A crucial part of the correct chiral behavior of the light pseudoscalar
meson octet in the DS approach, where the pseudoscalars are both
$q\bar q$ bound states and (almost-) Goldstone bosons of the dynamically
broken chiral symmetry of QCD, is the linear dependence of the squared
pseudoscalar masses on the current quark masses $m_q$ as the chiral limit
is approached
\begin{equation}
M_P^2 = M_{f f'}^2 =
{\rm const} \, ({m}_f + {m}_{f'}) \, .
\label{M2_prop_m}
\end{equation}
It is therefore interesting to investigate the validity of this
Gell-Mann--Oakes--Renner (GMOR)-type relationship
in the present approach in the vacuum and at finite temperatures.

\begin{figure}[!ht]
\includegraphics[width=.45\textwidth,angle=0]{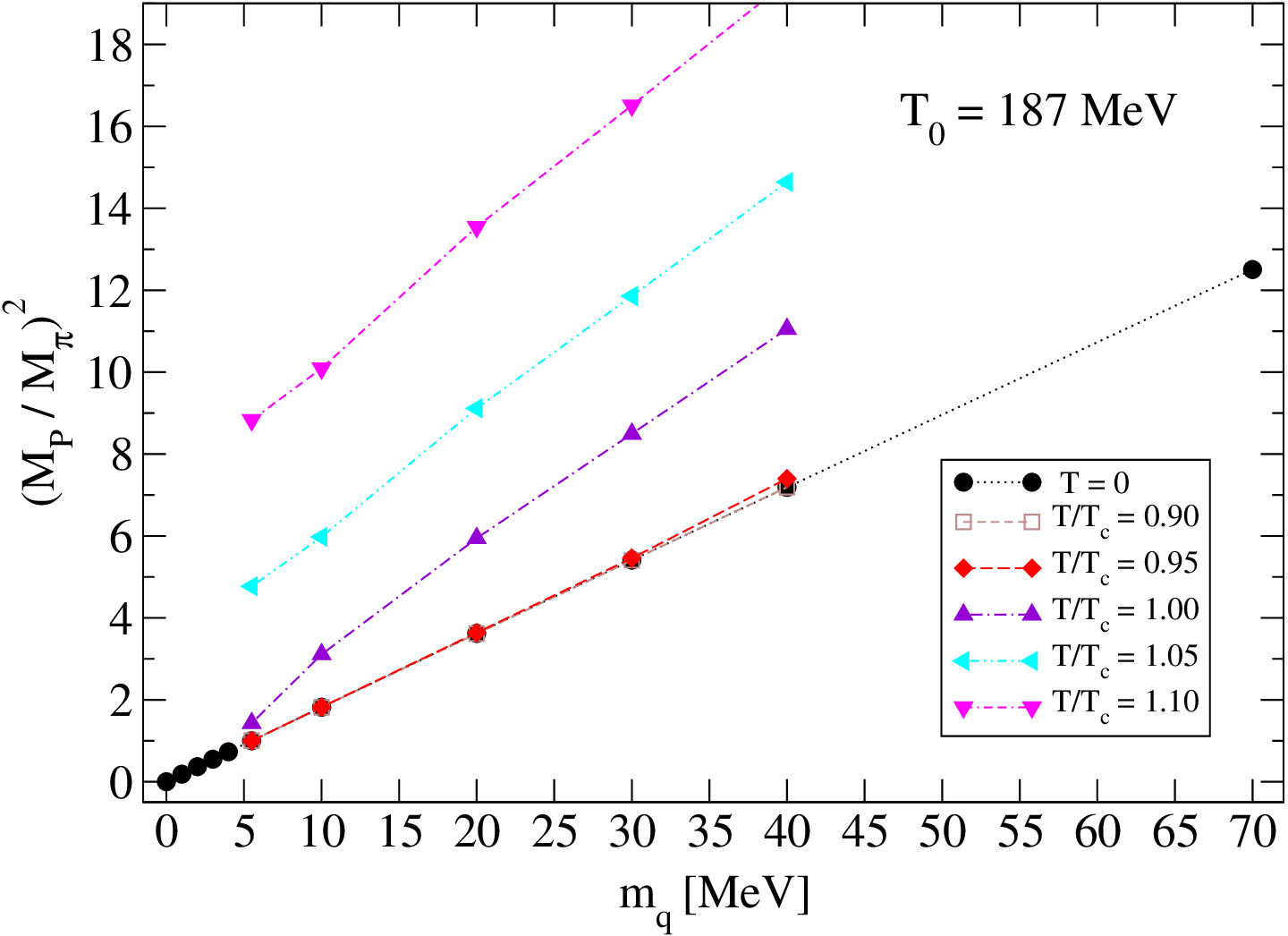}
\caption{
Squared pseudoscalar mass vs. current quark mass for different temperatures
(data symbols, connected by lines to guide the eye), show a GMOR-like 
behavior which holds in a wide range of quark masses, and up to temperatures 
very close to the critical one.
}
\label{GMOR-T}
\end{figure}

To this end, one calculates $M_P^2 = M_{f f'}^2$, the variable ``pion'' mass,
for different values of current light quark mass, $m_q$, while the current 
strange quark mass is kept fixed.
In Fig.~\ref{GMOR-T} we show that this relation is very well fulfilled
in the vacuum ($T=0$) up to current quark masses well exceeding $10~m_q^0$.
At finite temperatures, the GMOR-like relation (\ref{M2_prop_m})
qualitatively holds well up to $T\sim T_c$.

From the temperature independence of the pion mass up to $T_c$, together with
the validity of the GMOR-like relation in this range of temperatures, we can 
conclude that the temperature dependence of the chiral condensate must be 
mirrored by that of $f_\pi^2$. It is a question of utmost importance for 
the phenomenology of hadronic matter whether such a statement would also hold
when one goes beyond the rainbow-ladder level of description to which we 
restricted ourselves in the present work.

\section{Conclusions}

We have employed a Polyakov-loop  Dyson-Schwinger equation approach to 
investigate the pseudocritical temperatures for the chiral and 
deconfinement transitions for $N_f=2+1$ quark flavors using a 
rank-2 separable model for the effective gluon propagator. 
We find that the pseudocritical temperature  $T_\chi=128$ MeV for the chiral 
restoration and that for deconfinement, $T_d=270$ MeV, differ by more than a 
factor of two when the quark and gluon sectors are considered separately.
But  when the coupling is switched on these transitions get synchronized and 
the pseudocritical temperatures become coincident $T_c=T_\chi=T_d=195$ MeV.

We have investigated the dependence of $T_c$ on the parameter $T_0$ of the 
Polyakov-loop potential. For the logarithmic potential (\ref{logU}), we 
found two regions of linear dependence with a 
change in slope at $T_0=210$ MeV, accompanied with a change of the character
of the QCD transition from a crossover for $T_0>210$ MeV to a first oder
transition for $T_0<210$ MeV. It is a remarkable finding of the present work
that the $T_0$- rescaling to account for a quark flavor and chemical potential
dependence of the PL potential, which was suggested in  \cite{Schaefer:2007pw} 
and investigated in greater detail in \cite{Schaefer:2009ui} for different
parameterizations of the PL potential, when applied to the nonlocal separable 
PDSE model considered here, results in an obvious contradiction with 
LQCD concerning the character of the QCD transition. 
While in LQCD for $N_f=2+1$ the finite-T transition is a crossover, in
the present model with $T_0=187$ MeV it is a first order transition.
Nevertheless, for a different form of the PL potential, namely
the polynomial form (\ref{polyU}), we find that the QCD transition
remains crossover even for the smallest considered values of $T_0$.

As a consequence for possible phenomenological applications of the presented
approach we discussed that the coupling of the Polyakov-loop to the chiral 
quark dynamics narrows the temperature region in which chiral symmetry is 
approached.
Quantitative measures for this region are the $\sigma$-$\pi$ squared mass 
difference ($M_\sigma^2-M_\pi^2$) and the difference of 
temperatures for opening the $\sigma \to \bar{q} q$ decay channel 
($T_{\sigma-\bar{q}q}$) and the closing of the $\sigma \to 2\pi$ decay 
($T_{\sigma-2\pi}$).

The narrowness of the QCD transition region ($\Delta T/ T_c$) obtained from 
these measures is at the one-percent level and thus much too small for an 
adequate description of the QCD transition as obtained in recent LQCD
studies (see, e.g., Ref.~\cite{Cheng:2010fe}).

We conclude that the separable PDSE approach provides an essential 
improvement of the chiral quark dynamics in PNJL models and nonlocal PNJL 
models which use a rank-1 separable ansatz for the quark interaction kernel,
since it provides a running of both, the dynamical quark-mass function and 
the wave-function renormalization in close agreement with LQCD 
simulations of the quark propagator. It also provides the strong-coupling 
aspect of a dynamical confinement mechanism due to the absence of real quark 
mass poles.

However, the investigation of the temperature dependence of the chiral and
deconfinement order parameters characterizing the (pseudo-)critical
temperature and the width of the QCD transition reveals also some inadequate
aspects of the present level of description of this transition. 
%However, the investigation of the temperature dependence of the chiral and 
%deconfinement order parameters characterizing the (pseudo-)critical 
%temperature and the width of the QCD transition shows that the rainbow-ladder 
%level of description is insufficient to adequately describe the details of 
%this transition. 
The critical temperature is too high and the transition region
is too narrow when compared with LQCD results. 
A rescaling of the PL-potential results in a lower value for $T_c$, in 
accordance with recent LQCD results, but at the price of 
a narrowing of the QCD transition region, for the logarithmic PL
potential even changing the
character of the transition to a first order one, in striking contradiction    
with LQCD. 

We expect that going beyond the rainbow-ladder level by including hadronic 
fluctuations beyond the mean field 
\cite{Herbst:2010rf,Skokov:2010wb,Radzhabov:2010dd}
will entail an improvement of the approach. 
As has been demonstrated recently by including $\pi$ and $\sigma$ 
fluctuations in a consistent $1/N_c$ scheme 
\cite{Blaschke:2009qk,Radzhabov:2010dd}, going beyond the mean field will lead 
to a lowering of the chiral transition temperature.
The width of the transition region, however, appears as a sensitive 
constraint for the choice of an appropriate functional form of the PL potential.
Its possible dependence on the inclusion of hadronic correlations deserves
a detailed study.
We plan to extend our work in this direction.

\begin{acknowledgments}
We thank R. Alkofer, S. Beni\'{c}, M. Blank, K.A. Bugaev, H. Gies, 
Yu.L. Kalinovsky, A. Krasnigg, J.M. Pawlowski, 
A.E. Radzhabov, K. Redlich and B.-J. Schaefer for their discussions and 
comments to this paper.
D.H. is grateful for 
and acknowledges that these materials are based on work financed with 
support from the National Foundation for Science, Higher Education 
and Technological Development of the Republic of Croatia.
D.K. acknowledges the partial support of Abdus Salam ICTP.
D.H. and D.K. also acknowledge the support of the
project No. 119-0982930-1016 of the Ministry of Science, Education and Sports 
of Croatia.
D.B. was supported by the Russian Foundation for Basic 
Research under grant No. 08-02-01003-a and by the Polish Ministry of Science 
and Higher Education (MNiSW) under grant No. NN 202 2318 37.
This work was supported in part by CompStar, a Research Networking Programme 
of the European Science Foundation and by the Polish MNiSW grant 
"COMPSTAR-POL" No. 790/N-RNP-COMPSTAR/2010/0.
\end{acknowledgments}

\end{document}